\documentclass{article}
\usepackage[T1]{fontenc}
\usepackage[utf8]{inputenc}
\usepackage[]{ismir} %
\usepackage{amsmath,cite,url}
\usepackage{graphicx}
\usepackage{color}

\usepackage{xurl}



\usepackage{booktabs}   %
\usepackage{multirow}   %
\usepackage{subcaption}
\usepackage{amsmath}
\usepackage{amssymb}

\usepackage{algorithm}
\usepackage{algpseudocode}
\usepackage{listings}
\usepackage{xcolor}
\usepackage{makecell}

\title{Music-{JEPA}: Learning a World Model of Sound from Action}

\multauthor
  {Ziyu Wang$^{1,3}$ \hspace{1cm} Kun Fang$^{2,4}$ \hspace{1cm} Yann LeCun$^{1,5}$}
  {$^{1}$ Courant Institute, New York University\quad
   $^{2}$ Schulich School of Music, McGill University \\
   $^{3}$ Mohamed bin Zayed University of Artificial Intelligence (MBZUAI) \\
   $^{4}$ Centre for Interdisciplinary Research in Music Media and Technology (CIRMMT)\\
   $^{5}$ Advanced Machine Intelligence (AMI Labs)\\
   {\tt\small \{ziyu.wang, yann.lecun\}@nyu.edu, kun.fang@mail.mcgill.ca}}
   
\def\authorname{Z. Wang, K. Fang, and Y. LeCun}

\usepackage[bookmarks=false,pdfauthor={\authorname},pdfsubject={\pdfsubject},hidelinks]{hyperref}

\begin{document}

\maketitle

\begin{abstract}
Joint Embedding Predictive Architectures (JEPA) have recently emerged as a paradigm for learning world models by predicting latent representations, offering a promising direction for self-supervised learning. While initial attempts have applied JEPA to the music domain, it remains unclear how such frameworks can naturally support the formation of a world model for music. In this work, we propose to learn a world model of piano sound using JEPA by framing music as an action-conditioned system: the audio is treated as the state, and the pianoroll as the instrument action. Given a current audio state and an action, the model predicts the resulting future audio state, mirroring how humans learn musical sound through interaction. The model is trained in a fully offline setting using paired audio–pianoroll data, without environment interaction. Experiments show that the learned model captures the relationships between musical actions and their resulting sound. The resulting representations support downstream tasks, including beat tracking, composer identification, and key estimation, and enable piano transcription via planning, by searching for actions that best explain a target sound.\footnote{The demo page of this project can be accessed via: \url{https://zzwaang.github.io/music-jepa-demo/}. Code and pretrained models will be released soon.}

\end{abstract}

\section{Introduction}\label{sec:introduction}

While humans may be passively exposed to a wide range of sound signals, understanding music typically arises from active engagement. For example, learning piano music often involves interacting with the instrument and understanding how actions on the keyboard produce acoustic outcomes. Through such interaction, cognitive studies suggest that humans develop internal representations that enable them to anticipate and mentally simulate musical outcomes without direct auditory input~\cite{huron2006sweet, hohwy2013predictive}. This learning process naturally suggests a \textit{music world model}: a self-supervised framework that learns music representation by modeling the temporal dynamics between actions (e.g., playing an instrument) and resulting states (e.g., sound) in the latent space.

\begin{figure}[t]
  \centering
  \includegraphics[alt={ISMIR 2025 template example image},width=0.94\linewidth]{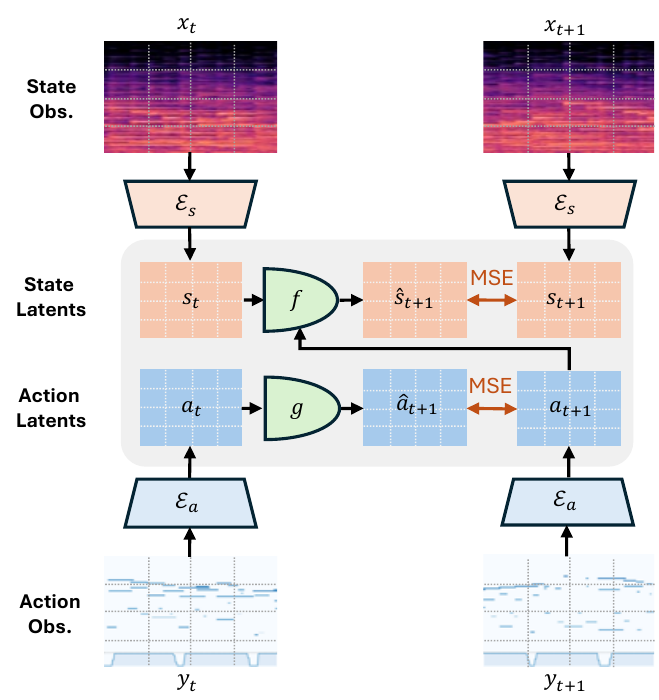}
  \caption{Model diagram of Music-JEPA. $\mathcal{E}_s$, $\mathcal{E}_a$, $f$, and $g$ denote the state encoder, action encoder, state predictor, and action predictor, respectively. The shaded region represents the latent temporal dynamics.}
  \label{fig:jepa_model}
\end{figure}

The music world model can be naturally instantiated within the Joint Embedding Predictive Architecture (JEPA)~\cite{LeCun2022APT}, which is framed to predict representations directly in a latent space without reconstructing raw observations, avoiding the burden of modeling unnecessary details. Originally developed in vision~\cite{assran2023self, assran2025vjepa2}, JEPA has only recently been explored in music information retrieval (MIR)~\cite{fei2023ajepa, tuncay2025audiojepa, riou2024investigating, riou2024stem}. Existing work primarily treats JEPA as a self-supervised approach for learning general audio representations in a passive way, as an alternative to contrastive learning~\cite{kong2025Emergent, spijkervet2021contrastive} or autoencoding~\cite{huang2022masked, li2024mert}. As a result, these approaches do not model how music evolves through time and are rarely evaluated on music-specific concepts such as key, chord, or beat. It remains unclear whether JEPA-based representations can capture temporal dynamics required for music understanding and planning.

In this work, we propose Music-JEPA, a music world model that captures latent temporal dynamics of audio conditioned on performance actions. As shown in Figure~\ref{fig:jepa_model}, we treat a 2-second piano audio segment as the state, and the corresponding pianoroll and sustain pedal signals as the action. The model predicts the next state from the previous state and current action directly in the latent space, without reconstructing raw audio. To prevent representation collapse, we adopt an exponential moving average (EMA) training scheme~\cite{Cai2021Ema}. Although trained offline without interaction, the learned dynamics capture coherent relationships between states and actions, including pitch, timing, velocity, and pedal control, and outperform JEPA models trained on passive audio alone.

We further evaluate Music-JEPA on downstream MIR tasks, including beat tracking, composer identification, and key detection on classical piano datasets. The learned representations consistently outperform the audio-only JEPA baseline and achieve performance comparable to MERT, despite using only 7\% of its parameters. We also explore planning using the learned dynamics via amortized planning~\cite{marion2021iterative}. Specifically, we train an inverse predictor to infer action representations and generate outputs with a separately trained decoder, enabling piano transcription through planning. The resulting transcriptions are coherent and perceptually meaningful, and are particularly effective at capturing continuous pedal variations, which remain challenging for supervised methods~\cite{zhang2025high}. Overall, we show that Music-JEPA provides a new learning framework that can understand music, predict the future, and plan.

\section{Related Works}\label{sec:relatedwork}

Learning meaningful representations is a fundamental problem not only in music but more broadly in machine intelligence. While recent advances in generative modeling have achieved strong performance, they often emphasize reproducing statistical patterns rather than capturing the underlying structure of how signals evolve~\cite{xing2025critiques}. This limitation has motivated the study of world models, where the goal is to learn representations that reflect system dynamics over time, rather than merely predicting observations~\cite{xiang2025pan, huang2026pointworld}. Among these approaches, JEPA~\cite{LeCun2022APT} is of particular interest because it predicts future states directly in a latent space without reconstructing raw observations, which is advantageous for high-dimensional inputs that may contain unpredictable or irrelevant details. This predictive perspective aligns with predictive coding theories in cognitive science~\cite{friston2005theory, hohwy2013predictive}, and also with music cognition, where internal anticipation has been argued to shape listening experience~\cite{huron2006sweet}, pitch and tonal expectations~\cite{krumhansl2001}, and rhythmic or metric perception~\cite{london2012}.

JEPA is proposed in the context of self-supervised learning (SSL), where the goal is to learn informative latent representations while preventing collapse. In vision, this paradigm has led to effective frameworks for image and video representation learning, with demonstrated success in downstream tasks and planning~\cite{assran2023self, assran2025vjepa2}. Various techniques have been proposed to regularize the latent space, including variance-based regularization~\cite{zhu2023vcreg}, self-distillation approaches~\cite{caron2021emerging}, exponential moving average (EMA)~\cite{Cai2021Ema}, and more recent distributional constraints~\cite{balestriero2025lejepa}. Among these, EMA-based methods have emerged as simple and effective solutions and are adopted in this work.

Building on these developments, several self-supervised approaches have been explored in the MIR domain. Recent work adapts JEPA-style architectures to audio by predicting masked regions of spectrograms or modeling coherence between audio tracks~\cite{fei2023ajepa, tuncay2025audiojepa, riou2024investigating, riou2024stem}. Other approaches employ contrastive learning~\cite{spijkervet2021contrastive, wilkins2025balancing}, self-distillation~\cite{anton2023audio, kong2025Emergent}, masked prediction~\cite{huang2022masked, li2024mert}, or invariance-based objectives~\cite{riou2023pesto, wu2025unsupervised} to learn audio representations. While these methods achieve strong performance on general audio tasks, they primarily operate on passive observations and do not explicitly model how music evolves through time or how actions influence sound, leaving action-conditioned dynamics largely unexplored.

\section{Method}
Music-JEPA models music as a dynamical system, where audio evolves over time under the influence of performance actions. We instantiate this framework on piano music using paired audio and control sequences $(x_{1:T}, y_{1:T})$, where $x_t$ is a short audio segment and $y_t$ the corresponding pianoroll and pedal signals. In our implementation, each segment spans 2 seconds.

As shown in Figure~\ref{fig:jepa_model}, we learn latent state and action representations $s_t=\mathcal{E}_{\text{state}}(x_t)$ and $a_t=\mathcal{E}_{\text{action}}(y_t)$, and model the dynamics as:

\begin{equation}
   s_{t+1} = f(s_t, a_{t+1}), \quad a_{t+1} = g(a_t).
\end{equation}
Here, $f(\cdot,\cdot)$ predicts the next state, and $g(\cdot)$ models the temporal structure of actions. Unlike typical settings where actions are low-dimensional, piano control is high-dimensional and correlated over time; modeling its evolution provides a structured prior over the action space.\footnote{In standard formulations, transitions are written as $f(x_{t}, a_{t})$. Here, we use $f(x_t, a_{t+1})$ since actions are aligned with the audio segment they generate in time, i.e., $a_{t+1}$ produces $s_{t+1}$.}

\subsection{Data Representation}
We use paired, time-aligned piano audio and MIDI data, segmented into 2-second intervals. Audio is represented as log-mel spectrograms with $F=229$ mel bins at a 10\,ms frame rate, yielding $x_t \in \mathbb{R}^{L \times F}$ with $L=200$. We use $n_{\text{fft}}=2048$ and a hop length of 160.

The action $y_t$ is derived from MIDI and aligned to the same frame rate as the audio. It consists of two components: a pianoroll $y_t^{\text{note}} \in \mathbb{R}^{L \times P}$ with $P=88$ pitches, and a sustain pedal signal $y_t^{\text{pedal}} \in \mathbb{R}^{L}$. Each entry of the pianoroll encodes note velocity over time.

\subsection{Model Architecture}

Our model follows the Joint Embedding Predictive Architecture (JEPA) framework, using a Vision Transformer (ViT) backbone~\cite{dosovitskiy2021vit} to encode both state and action representations. We treat the spectrogram and pianoroll as image-like inputs and map them into patch-based representations.

The spectrogram $x_t \in \mathbb{R}^{L \times F}$ is divided into $K_s = K_T \times K_F$ patches along time and frequency. Each patch corresponds to a local region of size $(c_T, c_F) = \lceil \frac{L}{K_T}, \frac{F}{K_F} \rceil$ and is projected into a $D$-dimensional embedding via a convolutional layer. Similarly, the pianoroll $y_t^{\text{note}} \in \mathbb{R}^{L \times P}$ is partitioned into $K_a = K_T \times K_P$ patches along time and pitch, and embedded using another convolutional projection. The sustain pedal signal $y_t^{\text{pedal}} \in \mathbb{R}^{L}$ is treated as a one-dimensional temporal sequence, partitioned into $K_T$ segments and encoded with a 1D convolution. Its representation is broadcast across pitch patches and fused with the pianoroll embeddings. After patch embedding, the state and action tokens are processed by transformer encoders, producing latent representations:
\begin{equation}
s = \mathcal{E}_s(x_t) \in \mathbb{R}^{K_s \times D}, \quad 
a = \mathcal{E}_a(y_t) \in \mathbb{R}^{K_a \times D}.
\end{equation}

The state and action predictors are also ViT-based, with distinct designs for modeling their dynamics. The state predictor $f(s,a)$ operates on a partially masked state sequence, taking $s_{t}$ and a masked version of $s_{t+1}$ as input, and predicting the missing regions via self-attention. Cross-attention to the action tokens $a_{t+1}$ is applied at each layer to condition the prediction. In contrast, the action predictor $g(a)$ is a standard Transformer without cross-attention. It takes $a_{t}$ and the masked tokens at time $t+1$, and is trained to reconstruct $ a_{t} $.

\subsection{Training Objective and Regularization}  

Given inputs $(x_{t-1}, y_{t-1})$ and $(x_t, y_t)$, we encode their latent representations and perform prediction in the latent space. Let $\bm{\theta}=\{\theta_{f}, \theta_{g}, \theta_{\mathcal{E}_s}, \theta_{\mathcal{E}_a}\}$ denote the model parameters. The training objective is:
\begin{equation}\label{eq:loss}
    \mathcal{L}(\bm{\theta}) = \| f(s_t, a_{t+1}) - s_{t + 1} \|^2 + \lambda \| g(a_t) - a_{t + 1} \|^2,
\end{equation}
where $\lambda$ controls the strength of the action prior. Predicting in the latent space introduces the risk of representation collapse, where the encoder may converge to trivial constant outputs. To mitigate this, we adopt an exponential moving average (EMA) strategy~\cite{Cai2021Ema}, maintaining stop-gradient teacher encoders with parameters $\bar{\theta}$ updated as
\begin{equation}
\bar{\theta}_{\mathcal{E}_k} \leftarrow \tau \bar{\theta}_{\mathcal{E}_k} + (1 - \tau)\theta_{\mathcal{E}_k}, \quad k \in \{s, a\},
\end{equation}
where $\tau$ is a momentum coefficient. The teacher encoders produce the target representations $s_{t+1}$ and $a_{t+1}$ in Eq.~\ref{eq:loss}, providing stable training target.

In addition, the model may reduce the loss by simply increasing the magnitude of representations without improving their alignment. To prevent this, we apply layer normalization to both student and teacher encoder outputs to control representation scale. With these design choices, the overall training procedure is summarized in Algorithm~\ref{algo:training}.

\begin{figure*}[t]
    \centering
    
    \begin{minipage}[t]{0.5325\textwidth}
        \centering
        \includegraphics[width=\linewidth]{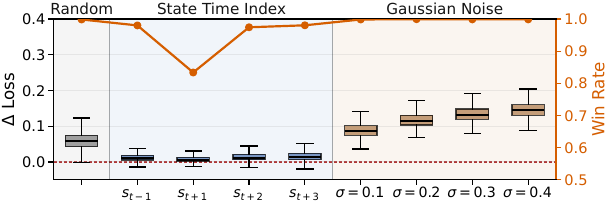}
        \subcaption{Input state perturbation.}
        \label{fig:expr_perturb:state}
    \end{minipage}
    \begin{minipage}[t]{0.3625\textwidth}
        \centering
        \includegraphics[width=\linewidth]{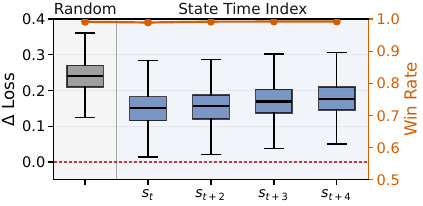}
        \subcaption{Target state perturbation.}
        \label{fig:expr_perturb:target}
    \end{minipage}

    \vspace{0.5em}

    \begin{minipage}[t]{0.91\textwidth}
        \centering
        \includegraphics[width=\linewidth]{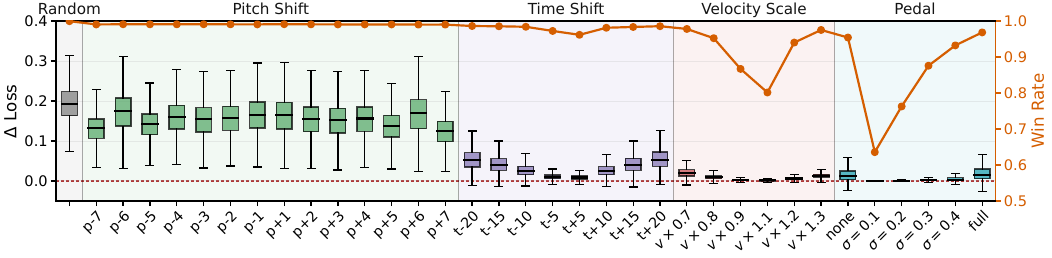}
    \vspace{-2em}
    \subcaption{Action perturbation.}
    \label{fig:expr_perturb:action}

    \end{minipage}
    
    \caption{Sensitivity to perturbations measured by the loss difference $\Delta \mathcal{L} = \mathcal{L}(\text{perturbed}) - \mathcal{L}(\text{correct})$ and win rate (fraction of cases with $\Delta \mathcal{L} > 0$). Top: input (left) and target (right) state perturbations via random sampling from the dataset, replacing segments at different time indices, and Gaussian noise on the input state with standard deviation proportional to spectrogram magnitude (e.g., 10\%, 20\%). Bottom: action perturbations including random actions, pitch shifts (semitones), temporal shifts (frames, 10\,ms), velocity scaling (scaling from 0.7-1.3), and pedal variations (removal, full activation, and additive noise with standard deviation proportional to velocity range).}

    \label{fig:expr_perturb}
\end{figure*}

\begin{figure*}
  \centering
  \includegraphics[alt={ISMIR 2025 template example image},width=0.96\linewidth]{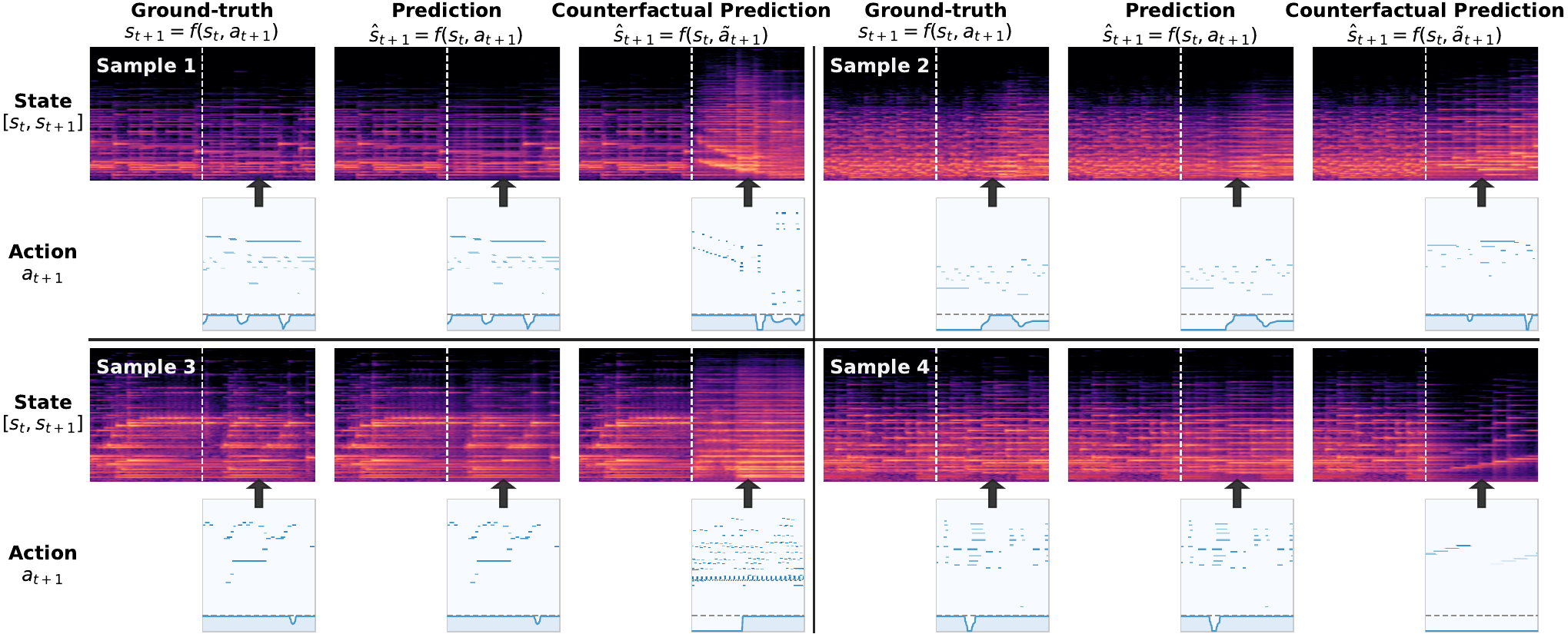}
    
  \caption{Four action-conditioned synthesis examples given $(s_t, a_{t+1})$ pairs from the test set. In each group: left, ground truth; middle, model prediction; right, counterfactual prediction with a different action.}
  \label{fig:synthesis}
\end{figure*}

\subsection{Downstream Tasks and Planning}\label{sec:method_planning}

After training, we freeze the model and use the state encoder as a feature extractor, producing representations of size $(K_s, D)$ for each 2-second audio segment. For longer audio of length $T$ seconds, representations are concatenated along the temporal dimension, resulting in a sequence of size $\left(\frac{T}{2} \times K_s, D\right)$. We evaluate the learned representations on beat tracking, composer identification, and key recognition in Section~\ref{sec:downstream}.

Beyond representation learning, the learned dynamics enable piano transcription via planning in latent space, i.e., inferring actions that produce the desired audio states. Given a known sequence of latent states $s_{1:T}$, the problem can be formulated as:
\begin{equation}
\small
a_{1:T}^* = \operatorname*{argmin}_{a_{1:T}} 
\sum_{t=1}^{T-1} \Big( 
\| f(s_t, a_{t+1}) - s_{t+1} \|^2 
+ \| g(a_t) - a_{t+1} \|^2 
\Big).
\end{equation}
In practice, the action space is high-dimensional, and gradient-based planning at this scale remains challenging~\cite{Parthasarathy2025Closing, wang2026temporal}. Unlike many planning settings where actions are low-dimensional control variables, Music-JEPA plans over dense latent action representations corresponding to expressive piano performances, resulting in a substantially larger optimization space. We therefore adopt an amortized optimization approach by training an inverse predictor $a_{t+1} = h(s_t, s_{t+1}, a_t)$ to approximate the solution~\cite{amos2022tutorial, nguyen2026latentgeometrysearchamortizing}. The predicted latent actions are then mapped back to the observation space using a separately trained action decoder. The JEPA parameters are kept frozen during the training of the inverse predictor and decoder. Experiment details are provided in Section~\ref{sec:transcription}.

\begin{algorithm}[tb]
   \caption{Pseudocode of Music-JEPA training.}
   \label{algo:training}
    \definecolor{codeblue}{rgb}{0.25,0.5,0.5}
    \lstset{
      basicstyle=\fontsize{7.2pt}{7.2pt}\ttfamily\bfseries,
      commentstyle=\fontsize{7.2pt}{7.2pt}\color{codeblue},
      keywordstyle=\fontsize{7.2pt}{7.2pt},
    }

\begin{lstlisting}[language=python]
# Es_stu, Es_tea: student and teacher state encoders
# Ea_stu, Ea_tea: student and teacher action encoders
# f, g: state and action predictors
# sg: stop-gradient operator; LN: layer normalization
# tau: EMA momentum; lambda: action pred. loss weight

Es_tea.params = Es_stu.params
Ea_tea.params = Ea_stu.params

for states, actions in loader:
    x_curr, x_next = states  # spectrogram (audio)
    y_curr, y_next = actions  # pianoroll + pedal

    # Student representations
    s_curr = LN(Es_stu(x_curr))
    a_curr = LN(Ea_stu(y_curr))
    a_next = LN(Ea_stu(y_next))

    # Teacher targets
    s_next_tgt = sg(LN(Es_tea(x_next)))
    a_next_tgt = sg(LN(Ea_tea(y_next)))

    # Prediction
    s_pred = f(s_curr, a_next)
    a_pred = g(a_curr)

    # Optimization
    loss = MSE(s_pred, s_next_tgt) + \\
        lambda * MSE(a_pred, a_next_tgt)
    loss.backward()
    
    # SGD/Adam update
    optimizer_step(Es_stu, Ea_stu, f, g)  

    # EMA update
    Es_tea.params = tau * Es_tea.params + \\
        (1 - tau) * Es_stu.params
    Ea_tea.params = tau * Ea_tea.params + \\
        (1 - tau) * Ea_stu.params
\end{lstlisting}
\end{algorithm}

\section{Experiments}\label{sec:experiment}
In this section, we evaluate Music-JEPA. We first describe the dataset, training procedure (Section~\ref{sec:dataset_training}), and baselines (Section~\ref{sec:baselines}). We then assess the model from three perspectives: temporal dynamics (Section~\ref{sec:temporal_dynamics}), representation quality on downstream MIR tasks (Section~\ref{sec:downstream}), and transcription via planning (Section~\ref{sec:transcription}).

\subsection{Dataset and Training Details}\label{sec:dataset_training}
We train Music-JEPA on the MAESTRO v3.0.0 dataset~\cite{hawthorne2018enabling}, which contains approximately 200 hours of piano recordings with time-aligned MIDI captured from Yamaha Disklavier pianos. The repertoire is primarily classical, spanning composers from the 17th to early 20th century. We use the standard train/validation/test splits. For the spectrogram, we use a patch size of $25 \times 15$, yielding $K_s = 16 \times 16 = 256$ patches. For the pianoroll, we use a patch size of $25 \times 6$, yielding $K_a = 16 \times 15 = 240$ patches. With embedding dimension $D=256$, the resulting state and action representations are of size $(256, 256)$ and $(240, 256)$, respectively. The state encoder uses 12 Transformer layers, and the action encoder uses 8 layers with the same architecture. Both the state and action predictors are 6-layer Transformers. All modules use $d_\text{model}=256$, $d_\text{ff}=512$, and 4 attention heads. The state encoder and predictor contain 6.5M and 4.9M parameters, while the action encoder and predictor contain 4.3M and 3.3M parameters, for a total of approximately 19M parameters.

We set $\lambda=0.5$ in Eq.~\ref{eq:loss} and the EMA momentum $\tau=0.95$. The model is trained with Adam~\cite{kingma2015adam} using a learning rate of $6\times10^{-4}$ and weight decay of $10^{-4}$, on a single NVIDIA A100 GPU with batch size 128. Since EMA-based training involves a moving target and the loss may not reflect convergence~\cite{Grill2020byol}, we empirically train for 15--25 epochs, which we find to yield stable results.

\subsection{Baseline Models}\label{sec:baselines}

Music-JEPA is evaluated against two types of baselines. First, to assess the benefit of action-conditioned dynamics, we compare against an existing JEPA that operates solely on audio via random patch masking without action conditioning (referred to as ``passive'' JEPA in Section~\ref{sec:introduction}). This baseline follows prior work such as Audio-JEPA~\cite{tuncay2025audiojepa} and A-JEPA~\cite{fei2023ajepa}. Since these models are trained on general audio and no public checkpoints are available, we re-implement and train this model on MAESTRO, denoted as \textbf{AO-JEPA} (audio-only JEPA). 

Second, we compare against \textbf{MERT}~\cite{li2024mert}, a large-scale pre-trained model widely used for music understanding, as a representative non-JEPA baseline. We also note that another recent self–distillation–based approach exists~\cite{kong2025Emergent}, but is not included as publicly available implementations or checkpoints are not available.

\begin{table}[t!]
\centering
\small
\setlength{\tabcolsep}{4pt}
\renewcommand{\arraystretch}{0.8}
\begin{tabular}{lcccc}
\toprule
\multirow{2}{*}{\textbf{Model}}  & \multicolumn{2}{c}{\textbf{Input State}} & \multicolumn{2}{c}{\textbf{Target State}} \\
\cmidrule(lr){2-3} \cmidrule(lr){4-5}
 & Temporal & Random & Temporal & Random \\
\midrule
Ours     & \textbf{0.929} & \textbf{0.999} & \textbf{0.991} & \textbf{0.992} \\
AO-JEPA  & 0.787 & 0.986 & 0.576 & 0.984 \\
\bottomrule
\end{tabular}

\caption{
Win rate under temporal and random perturbations of input and target states. Temporal win rates are averaged over all time-index shifts (Figure~\ref{fig:expr_perturb:state}~and~\ref{fig:expr_perturb:target}); random denotes samples drawn from the dataset. Higher values indicate stronger sensitivity to mismatched transitions.
}
\label{tab:win_rate}
\end{table}

\begin{table*}[h!]
\centering
{\small
\setlength{\tabcolsep}{3.2pt}
\renewcommand{\arraystretch}{0.88}
\begin{tabular}{l c cccccc cccc ccc}
\toprule
\multirow{2}{*}{\textbf{Model}} 
& \multirow{2}{*}[-0.25em]{\textbf{\shortstack{Encoder \\ Params.}}}
& \multicolumn{6}{c}{\textbf{Beat Tracking}}
& \multicolumn{4}{c}{\textbf{Composer Identification}}
& \multicolumn{3}{c}{\textbf{Key Recognition}} \\
\cmidrule(lr){3-8} \cmidrule(lr){9-12} \cmidrule(lr){13-15}
& & P$_{70}$ & R$_{70}$ & F1$_{70}$
& P$_{100}$ & R$_{100}$ & F1$_{100}$
& wF1 & Top-1 & Top-3 & Top-5
& wP & wR & wF1 \\
\midrule

\textbf{Ours}
& 6M
& \textbf{0.5346} & 0.8159 & \textbf{0.6208}
& \textbf{0.5695} & 0.8634 & \textbf{0.6599}
& 0.5520 & 0.3540 & \textbf{0.7045} & \textbf{0.8734}
& 0.7638 & \textbf{0.7625} & \textbf{0.7617} \\

AO-JEPA
& 6M
& 0.4929 & \textbf{0.8506} & 0.6013
& 0.5185 & \textbf{0.8941} & 0.6325
& 0.4606 & 0.3308 & 0.6635 & 0.8422
& \textbf{0.7657} & \textbf{0.7625} & 0.7615 \\

MERT
& 95M
& 0.4932 & 0.7368 & 0.5780
& 0.5379 & 0.8040 & 0.6309
& \textbf{0.6690} & \textbf{0.4059} & 0.6493 & 0.7996
& 0.6580 & 0.6485 & 0.6469 \\

\bottomrule
\end{tabular}
}
\caption{
Comparison of pretrained music representations on three downstream MIR tasks. Beat tracking is evaluated using precision, recall, and F1 at 70\,ms and 100\,ms onset tolerance thresholds. Composer identification is evaluated using weighted F1 (wF1) and Top-1/3/5 accuracy. Key recognition is evaluated using weighted precision (wP), recall (wR), and F1 (wF1). Higher is better for all metrics.
}
\label{tab:mir_tasks}
\end{table*}

\subsection{Evaluation of Latent Dynamics}\label{sec:temporal_dynamics}

We evaluate whether Music-JEPA captures action-conditioned temporal dynamics in the latent space. Given an input state $s_t$ and action $a_{t+1}$, the model predicts the next state $\hat{s}_{t+1} = f(s_t, a_{t+1})$, and we measure the error against the ground truth $s_{t+1}^{\text{gt}}$:
\begin{equation}
\mathcal{L}(s_t, a_{t+1}, s_{t+1}^{\text{gt}}) = \| f(s_t, a_{t+1}) - s_{t+1}^{\text{gt}} \|^2.
\end{equation}
A well-formed dynamics model assigns lower error to correct state–action–target triplets and higher error when any component is perturbed. Figure~\ref{fig:expr_perturb} summarizes the results. We visualize the distribution of loss differences $\Delta \mathcal{L}$ under perturbations relative to ground truth using box plots, where positive values indicate increased prediction error after perturbation. We also report the win rate, defined as the fraction of cases with $\Delta \mathcal{L} > 0$.

Figures~\ref{fig:expr_perturb:state} and~\ref{fig:expr_perturb:target} show perturbations to the input and target states, respectively. For the input state, we apply temporal shifts (replacing it with nearby segments) and additive Gaussian noise; for the target state, we apply temporal shifts. Both yield win rates close to 1.0, indicating strong sensitivity to mismatched states. When the input is replaced with $s_{t+1}$, the win rate is lowest (yet still $>0.8$), suggesting partial tolerance to repetition. The loss difference also increases with temporal distance, indicating sensitivity to temporal structure. Figure~\ref{fig:expr_perturb:action} shows action perturbations, including pitch shifts, temporal shifts, velocity scaling, and pedal variations. Pitch shifts have the strongest effect, particularly at tritone and other dissonant intervals. For other perturbations, sensitivity decreases for small temporal shifts (within 50\,ms) and minor velocity or pedal changes (e.g., 10\% scaling). Overall, the model responds consistently to both state and action perturbations.

Table~\ref{tab:win_rate} compares Music-JEPA with the AO-JEPA baseline using win rates under input and target state perturbations. For temporal shifts, we report win rates averaged across all time indices in Figures~\ref{fig:expr_perturb:state} and~\ref{fig:expr_perturb:target}. Without action conditioning, AO-JEPA captures temporal dynamics less effectively and shows limited ability to distinguish the correct next state from nearby alternatives (e.g., $s_{t+1}$ vs.\ $s_t$ or $s_{t+2}$). This suggests that while masked prediction captures temporal proximity, incorporating actions enables more accurate modeling of state transitions.

Finally, to further verify the learned dynamics, we freeze the encoder and train a spectrogram decoder to reconstruct $\hat{x}_{t+1}$ in the spectrogram domain. As shown in Figure~\ref{fig:synthesis}, with ground-truth actions, the model produces outputs that closely match the ground truth (first column). When given counterfactual actions (e.g., random pianoroll samples, arpeggios, or scales), the predicted spectrogram changes accordingly. These results suggest that the model captures informative and non-trivial action-conditioned dynamics through latent-space prediction alone.

\subsection{Evaluation on MIR Downstream Tasks}\label{sec:downstream}

We evaluate representations learned by the Music-JEPA state encoder on MIR tasks, including beat tracking, composer identification, and key recognition on classical piano music. We use 4-second clips for beat and key tasks, and 12-second clips for composer identification to capture longer-term structure. We compare against AO-JEPA and MERT, with results in Table~\ref{tab:mir_tasks}. Overall, Music-JEPA outperforms AO-JEPA on all tasks and remains competitive with MERT despite using only 7\% of its parameters. We do not claim to outperform MERT in general, as it is trained on broad audio and not specialized for classical piano.

\textbf{Beat tracking} We use the ASAP dataset~\cite{foscarin2020asap}, which provides a subset of MAESTRO audio and MIDI with beat annotations. For both JEPA-based models and MERT, we use a comparable probing head of approximately 4M parameters, consisting of four ViT blocks followed by convolutional layers for frame-level prediction. Training uses the loss from~\cite{foscarin2024beat} together with standard Madmom post-processing~\cite{madmom}. As shown in Table~\ref{tab:mir_tasks}, results are reported under 70\,ms and 100\,ms tolerance thresholds, with Music-JEPA achieving the best overall F1. Beat tracking in classical piano music remains particularly challenging due to weak percussive cues and rich temporal structure, and remains difficult even for supervised methods~\cite{foscarin2024beat}.

\begin{figure}[h]
    \centering
    \begin{subfigure}{0.48\linewidth}
        \centering
        \includegraphics[width=\linewidth]{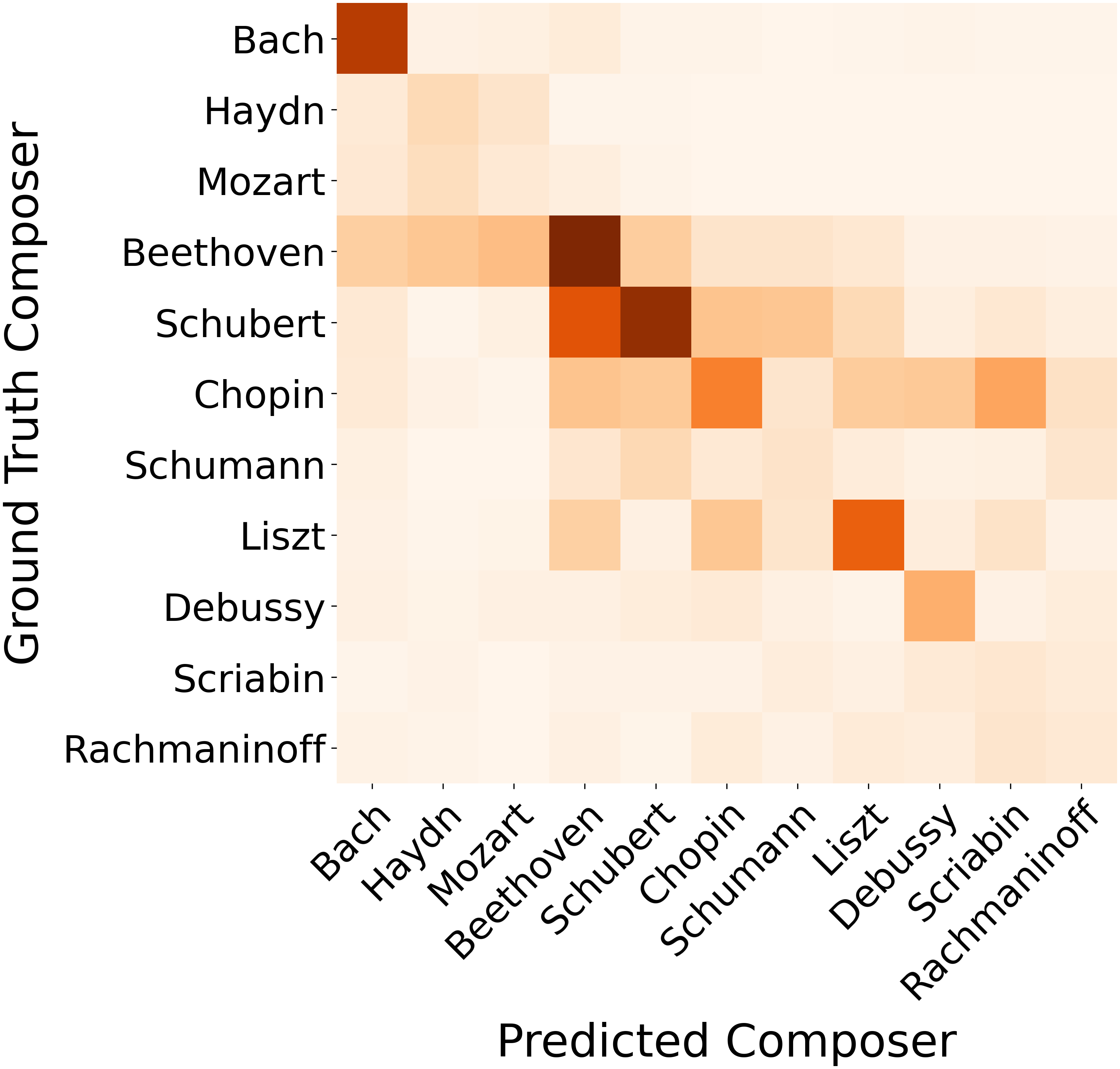}
        \caption{Our model.}
    \end{subfigure}
    \hfill
    \begin{subfigure}{0.44\linewidth}
        \centering
        \includegraphics[width=\linewidth]{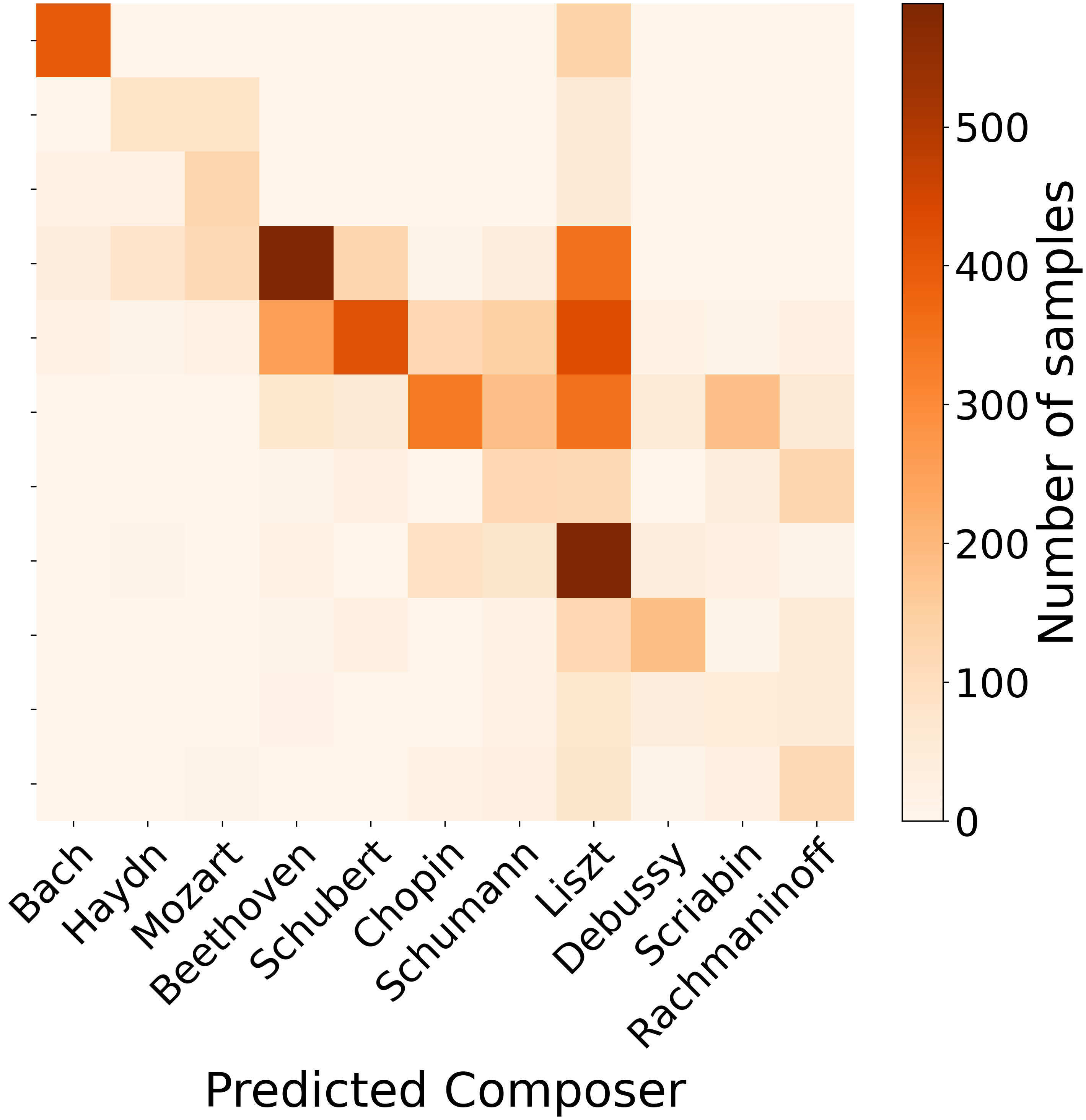}
        \caption{MERT.}
    \end{subfigure}
    \caption{Comparison of confusion matrices for classical composer identification task.}
    \label{fig:composer_conf_mat}
\end{figure}

\textbf{Composer identification} We train a downstream classifier using composer labels from MAESTRO. We remove pieces with multiple composer annotations (e.g., Liszt arrangements of Beethoven) and retain the 11 most frequent composers appearing in both the training and test sets. The probe is a 2-layer MLP with $\sim$16K parameters that aggregates frequency patches (or MERT latent dimensions), followed by mean pooling over time. Table~\ref{tab:mir_tasks} shows that Music-JEPA outperforms AO-JEPA and achieves the highest Top-3 and Top-5 accuracy, though its Top-1 accuracy and weighted F1 are lower than MERT. Figure~\ref{fig:composer_conf_mat} shows that Music-JEPA yields structured confusions aligned with music history, such as between Haydn and Mozart and among Romantic composers, while Beethoven bridges Classical and Romantic styles. MERT, in contrast, makes less structured errors and tends to over-predict certain composers (e.g., Liszt), suggesting less interpretable behavior.

\textbf{Key recognition} Although ASAP provides key signature annotations, modulations within a piece without changing the key signature are common in classical music. To assess how scale-related concepts are represented, we construct pseudo key labels by matching pianoroll chroma features to the 12 major scales and selecting the closest match~\cite{krumhansl2001}. We use the same probe architecture as in composer identification. Table~\ref{tab:mir_tasks} shows that both JEPA methods outperform MERT by about 10 points across metrics, indicating that scale-related information is well captured.

\subsection{Transcription via Planning}\label{sec:transcription}

Finally, we explore leveraging the learned dynamics for piano transcription via planning, as formulated in Section~\ref{sec:method_planning}. Due to instability in high-dimensional gradient-based optimization, we adopt an amortized approach by training an inverse predictor $\hat{a}_{t+1}=h(s_t, s_{t+1}, a_t)$ to infer actions, analogous to policy learning in reinforcement learning. The predicted actions are then decoded into pianoroll representations using a separately trained decoder.

\begin{table}[h]
\centering
\small
\setlength{\tabcolsep}{4pt}
\begin{tabular}{lccccc}
\toprule
\multirow{2}{*}{\textbf{Model}} 
& \multicolumn{4}{c}{\textbf{Note F1}} 
& \textbf{Pedal} \\
\cmidrule(lr){2-5} \cmidrule(lr){6-6}
& \textbf{Frm.}$\uparrow$ & \textbf{Ons.}$\uparrow$ & \textbf{+Off.}$\uparrow$ & \textbf{+Vel.}$\uparrow$ & \textbf{MAE}$\downarrow$ \\
\midrule
Ours & 0.8381 & 0.7325 & 0.5372 & 0.4000 & \textbf{0.1222} \\
Kong et al.~\cite{kong2021high} & 0.9113 & 0.9681 & 0.8524 & 0.8377 & 0.1471 \\
Yan et al.~\cite{yan2024scoring} & \textbf{0.9375 }& \textbf{0.9833} & \textbf{0.9166} & \textbf{0.9122} & 0.1414 \\
Zhang et al.~\cite{zhang2025high} & --- & --- & --- & --- & 0.1339 \\
\bottomrule
\end{tabular}

\caption{Transcription and continuous sustain pedal estimation results on the MAESTRO test split. Note-level metrics include frame, onset, onset+offset, and onset+offset+velocity F1 scores. Arrows indicate whether higher or lower values are better. ``---'' indicates N/A.}
\label{tab:transcription}
\end{table}

We compare our transcription results with state-of-the-art methods, including Yan et al.~\cite{yan2024scoring} and Kong et al.~\cite{kong2021high}, as well as a supervised sustain pedal model by Zhang et al.~\cite{zhang2025high}. As shown in Table~\ref{tab:transcription}, while our model produces perceptually coherent transcriptions, its note-level performance remains below fully supervised methods. Following Zhang et al.~\cite{zhang2025high}, sustain pedal estimation is evaluated using the mean absolute error (MAE) between the predicted and ground-truth continuous pedal curves at 100 frames per second. In contrast, Music-JEPA achieves the best performance on continuous sustain pedal estimation. This may be because pedal control has a strong and direct impact on the resulting audio, making it more accessible to modeling via action-conditioned dynamics. By comparison, supervised approaches often rely on matching ground-truth pedal values, which may admit multiple equivalent patterns. Overall, these results highlight both the potential of planning-based transcription and the need for improved perceptual metrics and more stable planning strategies.

\section{Conclusion and Future Work}\label{sec:conclusion}
We propose Music-JEPA, a music world model that learns action-conditioned temporal dynamics of audio in latent space. We demonstrate three key capabilities. First, the model captures temporal dynamics more effectively than audio-only JEPA baselines, showing consistent sensitivity to state–action transitions. Second, the learned representations are informative, achieving competitive performance on piano MIR tasks compared to larger self-supervised models. Third, the learned dynamics enable planning, yielding meaningful transcription behavior. At the same time, it remains unclear how well the model scales or generalizes to broader musical settings where action annotations are unavailable. Extending the framework to more abstract or latent actions, and enabling more meaningful planning tasks such as composition, remain open challenges. We hope this work provides a foundation for future exploration.

\bibliography{ISMIRtemplate}

\end{document}